\documentclass[conference]{IEEEtran}
\IEEEoverridecommandlockouts
\usepackage{cite}
\usepackage{amsmath,amssymb,amsfonts}
\usepackage{algorithmic}
\usepackage{graphicx}
\usepackage{textcomp}
\usepackage{xcolor}
\usepackage{float}
\usepackage{gensymb}
\usepackage{siunitx}

\DeclareMathOperator*{\argmin}{argmin}

\def\BibTeX{{\rm B\kern-.05em{\sc i\kern-.025em b}\kern-.08em
    T\kern-.1667em\lower.7ex\hbox{E}\kern-.125emX}}

\makeatletter

\def\ps@IEEEtitlepagestyle{%
  \def\@oddfoot{\mycopyrightnotice}%
  \def\@evenfoot{}%
}
\def\mycopyrightnotice{%
  {\footnotesize 978-1-7281-9556-8/21/\$31.00 \textcopyright2021 IEEE \hfill}% <--- Change here depending on the required notice
  \gdef\mycopyrightnotice{}% just in case
}

\begin{document}
\title{SREDS: A dichromatic separation based measure of skin color\\
%\title{Skin Reflectance Estimate based on Dichromatic Separation (SREDS)\\
\thanks{This material is based upon work supported by the Center for Identification Technology Research and the National Science Foundation under Grant No.$650503$.}
}

\author{\IEEEauthorblockN{Keivan Bahmani}
\IEEEauthorblockA{\textit{Clarkson University} \\
Potsdam, NY, USA \\
bahmank@clarkson.edu}
\and
\IEEEauthorblockN{Richard Plesh}
\IEEEauthorblockA{\textit{Clarkson University} \\
Potsdam, NY, USA \\
pleshro@clarkson.edu}
\and
\IEEEauthorblockN{Chinmay Sahu}
\IEEEauthorblockA{\textit{Clarkson University} \\
Potsdam, NY, USA \\
sahuc@clarkson.edu}
\and
\IEEEauthorblockN{Mahesh Banavar}
\IEEEauthorblockA{\textit{Clarkson University} \\
Potsdam, NY, USA \\
mbanavar@clarkson.edu}\\
\and
\IEEEauthorblockN{Stephanie Schuckers}
\IEEEauthorblockA{\textit{Clarkson University} \\
Potsdam, NY, USA \\
sschucke@clarkson.edu}
}

\maketitle

\begin{abstract}
Face recognition (FR) systems are fast becoming ubiquitous. However, differential performance among certain demographics was identified in several widely used FR models. The skin tone of the subject is an important factor in addressing the differential performance. Previous work has used modeling methods to propose skin tone measures of subjects across different illuminations or utilized subjective labels of skin color and demographic information. However, such models heavily rely on consistent background and lighting for calibration, or utilize labeled datasets, which are time-consuming to generate or are unavailable. In this work, we have developed a novel and data-driven skin color measure capable of accurately representing subjects' skin tone from a single image, without requiring a consistent background or illumination. Our measure leverages the dichromatic reflection model in RGB space to decompose skin patches into diffuse and specular bases. 
\end{abstract}


\section{Introduction}
\label{intro}
\begin{figure}[htb]
\begin{minipage}[b]{1\linewidth}
  \centering
  \centerline{\includegraphics[width=11cm]{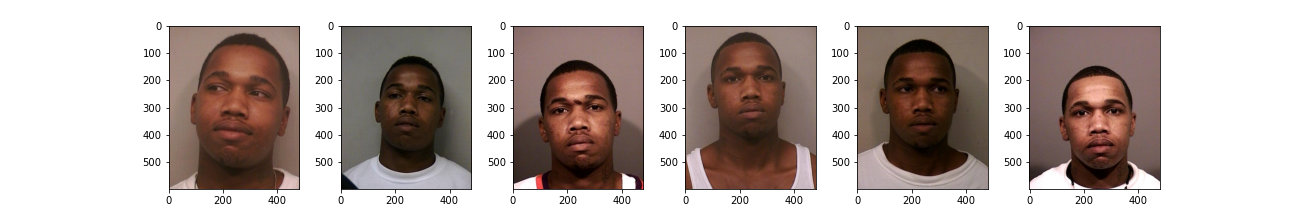}}
\end{minipage}
\caption{Variability of skin color of the same subject from MEDS-II dataset}
\label{fig:sample}
\end{figure}

Interest in facial recognition has been increasing rapidly as the technology has improved in performance and reliability over the past few decades. Facial recognition systems are commonly used in video authentication, criminal identification and building/device access control and many other areas \cite{kortli_face_2020}. Since facial recognition is involved in such critical applications, researchers are investigating how error rates differ between different demographic groups.
%In response to the proliferation of facial recognition technology into our society, scholars are raising alarm bells on the broad unintended consequences of the technology and have called for further study into their behavior \cite{rahwan_machine_2019}. 
A report by the National Institute of Standards and Technology (NIST) studied this question and
%e performance variation in facial recognition algorithms across different demographic groups,
found evidence of demographic differentials in the majority of algorithms evaluated \cite{grother_face_2019}. While the best algorithms did not present a differential, there is a desire to minimize this for all algorithms. Performance differential commonly comes about by maximizing overall predictive accuracy without considering how one subgroup's performance masks another's deficiencies \cite{madras_learning_2018}. 
%Research has emerged suggesting that the practice of maximizing overall predictive accuracy on real-world datasets leads to bias as one subgroup's performance often masks another's deficiencies \cite{madras_learning_2018}. 
Recent research has focused on providing solutions to differential performance in facial recognition, or often called "bias" by the popular press. The problem has two components, false negatives and false positives. Since each is a unique problem and bias encompasses a larger, diverse set of issues, biometric researchers prefer the term differential performance rather then bias \cite{howard_effect_2019}. The solutions typically fall into one of three categories: re-balanced training sets \cite{wang_mitigating_2020}, protected attribute suppression \cite{morales_sensitivenets_2020}, and model adaption \cite{wang_racial_2019}. Nearly all of these solutions require datasets containing demographic labels for applying their strategies and evaluating performance. While datasets like this exist \cite{founds2011nist,ricanek2006morph}, 
%,sixta2020fairface,wang_mitigating_2020
they represent a small fraction of the total number of datasets for facial recognition. Furthermore, datasets are often labeled for a single task \cite{alvi_turning_2019}, and demographic data may be overlooked because it is difficult to collect reliably. 
%This is because the information is often subjective and self-reported, leading to inconsistent definitions of what constitutes certain demographic groups.
For this reason, many researchers apply demographic information to datasets after collection. One approach is to use an off-the-shelf deep-learning-based ethnicity classifier. The authors in \cite{wang_mitigating_2020} utilize the proprietary Face++ API for labeling their datasets. While this method scales easily to large datasets, the failure points of these models are not well understood and the %highly non-linear and 
complex relationships they rely on are very difficult to interpret. Due to this limitation, many researchers rely on more readily apparent attributes such as skin color for labeling their datasets. While skin color doesn't directly represent demographic information, it does have correlation with ethnic self-definitions \cite{Jablonski_skin_color_race}. Nevertheless, determining a person's inherent skin color from an image can be challenging due to a person's natural skin variability, the camera parameters, and changes in lighting. Figure \ref{fig:sample} depicts the variation of skin tone for a subject in MEDS-II dataset \cite{founds2011nist}. A widespread method for estimating skin pigment from facial images is via Fitzpatrick skin type (FST) \cite{fitzpatrick1988validity}.
%, krishnapriya_issues_2020}. 
%This measure is a skin pigment estimator that has found widespread use in labeling facial images \cite{krishnapriya_issues_2020}. 
Despite its popularity, evidence is beginning to mount that FST has limited quantification and reliability \cite{rampen_unreliability_1988}, particularly for non-white individuals \cite{bino_variations_2013}.
%explain the change in skin tone can be natural (different seasons cite Geographic distribution of environmental factors influencing human skin coloration, duo to lighting and imaging equipment (various gains).  This results in high intra-subject skin tone variability cite john howards nist presentation.
In response to FST's limitations and off-the-self deep algorithm's poor interpretability, researchers have developed the skin color metrics Individual Typology Angle (ITA) \cite{chardon1991skin} and Relative Skin Reflectance (RSR)  \cite{cook_demographic_2019}.
%as metrics for detecting a subject's skin color from an image. 
 %interpretable models for estimating demographic data such as Individual Typology Angle (ITA)  \cite{del2006relationship} or Relative Skin Reflectance (RSR)  \cite{cook_demographic_2019}. These two methods generate a measure of skin tone that can be applied to datasets for skin color bias mitigation. 
%Our contribution in this work is a novel metric with superior performance with less restrictive assumptions required.
ITA utilizes colorimetric parameters to provide a point-wise estimate of the skin color represented in the image of a person. On the other hand, RSR is a data-driven approach that utilizes the distribution of skin pixels in color space to fit a linear model estimating skin tone. Both methods are sensitive to changes in illumination and RSR in particular requires a highly controlled acquisition environment (constant background, lighting and camera). These restrictions greatly hamper the usefulness of these metrics in the more challenging deployments of facial recognition without control over the collection environment.

%Since the method Even so, the method is sensitive to changes in illumination, and  this method is not robust to all types of noise in the image and provides no ability to fit to a particular dataset. RSR introduces a data driven approach that provides a continuous estimate of skin tone based on physical skin properties. However, RSR requires a controlled acquisition environment (constant background, lighting and camera). The xxx limitation of ITA makes it difficult for researchers. The highly constranied collection environment needed for RSR to work accurately removes the majority of facial datasets from consideration.

%\textcolor{red}{How do these shortcomings affect the research? Insert, here, a bridge sentence or two that will also serve to provide motivation for SREDS.} \textcolor{blue}{Does this work?} \textcolor{orange}{I took another crack at it. thoughts?}

Motivated by these shortcomings, we introduce Skin Reflectance Estimate based on Dichromatic Separation (SREDS). SREDS provides a continuous skin tone estimate by leveraging the dichromatic reflection model \cite{shafer_using_1985} and explicitly considering the different types of illumination across the face. This provides SREDS with superior or comparable performance in both less consistent and highly controlled acquisition environments. Additionally, the dichromatic model provides us greater interpretabilty into the locations of the face most utilized in the final metric generation.
%with specular and diffuese light decompositions to interpret the pixels  
%This allows the metric to require a less consistent and controlled acquisition environment, an interpretable separation of which face pixels are are representative of the skin color.  
%By decomposing the image into specular and diffue components, the skin pixels the mode us  

To evaluate this measure, we consider both its stability and meaningfulness compared to ITA and RSR over three different datasets: Multi-PIE Multi-view dataset \cite{sim2002cmu}, the Multiple Encounter Dataset (MEDS-II) \cite{founds2011nist}, and the Morph dataset \cite{ricanek2006morph}. We measure consistency using intra-subject variation and analyze meaningfulness by examining the distribution of ethnicities with respect to skin color estimates.
%Our results show that SREDS provides a more consistent estimate of skin tone (lower intra-subject variability) when compared to ITA and RSR while still providing a meaningful and interpretable measure.
Since bias mitigation frameworks optimize to reduce differential performance with respect to their demographic information, we expect any performance improvement in the demographic labeling to translate to more fair algorithms for end users. The next sections provide a description of skin metrics, preprocessing, datasets used, results, discussion, and conclusions.

%The rest of this is paper is organized as follows. The existing skin color metrics are presented in section \ref{sec:Benchmark}. Then, the bidirectional reflectance distribution function (BRDF) model is presented \ref{sec:HSM} and the derivation of the ``Skin Reflectance Estimate based on  Dichromatic  Separation''  (SREDS) metric is discussed in section \ref{sec:SREDS}. In section \ref{ssec:ROI}, the region of interest selection and the dataset is discussed. Then results are discussed in section \ref{sec:ResDis} and concluding remarks are drawn in section \ref{sec:conclusion}.

%We initially provide a brief overview of the dichromatic reflection model in section \ref{sec:Dichromatic Reflection}. Section \ref{sec:SREDS} describes the detail of SREDS. In Section ref{} we provide details of our evalauiton process and Finally ....

\section{Benchmark Skin Color Metrics}
\label{sec:Benchmark}
% As most of the face region is dominated by skin, 

%The appearance and perception of the color of a person are heavily influenced by the characteristics of the skin. Earlier studies investigated different characteristics of the skin based on skin color \cite{chardon1991skin,takiwaki1998measurement, wang2015measuring}, skin type \cite{chardon1991skin,fitzpatrick1988validity}, and skin reflectance \cite{weyrich2006analysis}.  Initial studies relied on  Fitzpatrick  skin  type  (FST) measure to classify sun-reactive skin types \cite{fitzpatrick1988validity}. However, there is no universal measure for skin color estimation. In what follows, we discuss two existing metrics, individual typology angle (ITA) and relative skin reflectance (RSR), that represent skin tone.

In this section we discuss our implementation of two existing skin color metrics, ITA and RSR. Each method has a unique pipeline from facial image to final metric, and we replicated the original work to the best of our ability.

% \textcolor{blue}{A sentence or two here introducing this section.}

\subsection{Individual Typology Angle}
\label{ssec:ITA}
% As most of the face region is dominated by skin, the appearance and perception of the color of a person are heavily influenced by the characteristic of the skin. Earlier studies investigated different characteristics of the skin based on skin color \cite{chardon1991skin,takiwaki1998measurement,wang2015measuring}, skin type \cite{chardon1991skin,fitzpatrick1988validity}, and skin reflectance \cite{weyrich2006analysis}.  Initial studies relied on Fitzpatrick skin type (FST) measure to classify sun-reactive skin types \cite{fitzpatrick1988validity}. However, even today there is no universal measure for skin color estimation.  

%Individual typology angle (ITA) is a tool for estimating the effects of ultraviolet radiation (UVR) exposure on skin. The ITA value provides an objective skin color type classification that is potentially predictive of the biological consequences of UVR exposure. In \cite{wilkes2015fitzpatrick}, authors investigated and found a high correlation between Melanin Index (which is used to assign FST types) and Individual Typology Angle (ITA) \cite{del2006relationship}. Therefore, ITA has been suggested as a reliable proxy from FST, automating the process and eliminating the need for a subject-matter expert. 
%Hence, without knowledge of FST and help from a dermatologist, the ITA can be estimated from an image. 

Individual typology angle (ITA) is a type of colorimetric analysis designed to measure acquired tanning \cite{chardon1991skin}. Given the simplicity of ITA and its correlation with Melanin Index \cite{merler2019diversity}, it is an ideal candidate for determining skin tone directly from an image. 
% To benchmark our proposed algorithm, we implemented the ITA algorithm \cite{merler2019diversity}. 
An RGB image is converted into CIE-Lab space \cite{connolly1997study}, as follows: (1) the `L' component which quantifies luminance, (2) the `a' component - absence or presence of redness, and (3) the `b' component - yellowness. Using the `L' and `b' components, Pixel-wise ITA value, in degrees, can be estimated throughout an image as: 
\begin{equation}\label{pixel-wise-ita}
    ITA = \frac{\arctan(L-50)}{b}*\frac{180}{\pi}. 
\end{equation}

In order to find suitable skin pixels in the image, a Dlib landmark extractor (see Section \ref{ssec:ROI}) is used to detect the forehead, left cheek, and right cheek facial regions. 
%We left out the chin region due to its small skin region and additional variability from facial hair. 
For each facial region, pixel-wise ITA is computed and smoothed using an averaging filter. The mode of each region's resulting distribution are averaged together to create a single skin tone estimate for a face.
%To generate a single ITA measure for a face, we first apply the pixel-wise ITA calculation to   from   patches such as right cheek, left cheek, and forehead are extracted using a Dlib landmark extractor (see Section \ref{ssec:ROI}). 
%To generate a single ITA value from an image, facial patches for face and cheeks are extracted 
%ITA values for each skin patch are estimated, fit to a Gaussian function, and the peak value extracted as the ITA score of the corresponding patch. The mean of ITA values of the patches are estimated to generate a fused single estimate of skin tone. 
%\textcolor{blue}{Details here, please.}

\subsection{Relative Skin Reflectance}
\label{ssec:Relative Skin Reflectance}
\begin{figure}[t]
\begin{minipage}[b]{1.0\linewidth}
  \centering
  \centerline{\includegraphics[width=9cm]{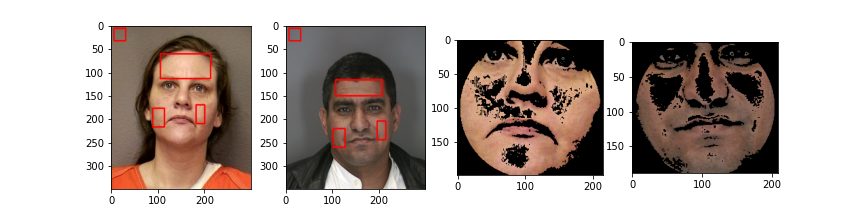}}
\end{minipage}
\caption{Region of interest extraction for cheeks,forehead, and background crops (Left). Adaptive skin segmentation (Right)}
\label{fig:ASS}
\end{figure}

 %Research has shown that skin pixel intensity in a digital image is affected by three factors: physical properties of the skin and underlying tissues, physical properties of the skin's surface, and imaging artifacts. 
 Relative Skin Reflectance (RSR) is a process designed to relate physical properties of the skin to the performance of facial recognition ~\cite{cook_demographic_2019}.
 The pipeline works by removing the confounding effects of imaging artifacts on skin pixels and then fitting a line in the direction of greatest variance in RGB color-space. 
 The resulting metric is related to the skin tone of each subject relative to the rest of the photos in the dataset.
% measure skin tone intensity, a value related to the tone of the skin, using the physical skin properties rather then imaging artifacts. 
 %``skin color.'' ~\cite{cook_demographic_2019}. The resulting metric is strongly related to net skin reflectance, a value related to skin tone of each subject relative to the rest of the enrollment photos in the study.
 %The values produced provide a metric that is strongly related to net skin reflectance of each subject relative to the rest of the enrollment photos in the study. 
 The computation of RSR begins with the selection of facial skin pixels using face detection, circular masking, and luminance outlier removal as described in \cite{taylor_adaptive_2014}. Pixel intensities are then corrected using divisive normalization via background sampling. We refer to this preprocessing pipeline as \emph{adaptive skin segmentation}. Figure \ref{fig:ASS} depicts the outcome of adaptive skin segmentation algorithm applied to two faces. After removal of non-skin and outlier pixels, the researchers assume that the direction of greatest variation in RGB color-space represents net skin reflectance. Linear principal component analysis (PCA) is used to fit a line in this direction.
 
%A final metric is produced by averaging the projection of skin pixels onto the first principal component. 
The final RSR metric is produced by averaging the projection of skin pixels onto the first principal component. Assumptions include consistent lighting, the same acquisition camera, and constant background. The Multi-PIE dataset was able to meet all conditions. However, due to the lack of constant background in MEDS-II and MORPH-II, the background normalization step was bypassed for these datasets. As a further limitation, the metric only gives an indication of where a subject lies in terms of net skin reflectance relative to the other subjects in the dataset, rather then an absolute measure.

 %, thereby creating a projection that is subsequently used for determining the relative skin reflectance of skin pixels in that dataset. 
 %The Multi-PIE dataset was able to meet all conditions. 
 %A final metric is produced by averaging together the projection of skin pixels on principal component one across a facial image.
 %when applied to the face image presented in Figure \ref{fig:compare}. 
 %This methodology has a number of assumptions that must be met from the image collection. 

\section{Human Skin Analysis}
\label{sec:HSM}
%Previous work on reflectance modeling of human faces relied on heavily controlled and calibrated image acquisition, \textit{i.e.}, acquiring multiple samples under controlled illumination or with different polarization. The acquired data is fit to bidirectional reflectance distribution function (BRDF) and bidirectional surface scattering distribution function (BSSRDF) models \cite{jensen2001practical} to fully represent the spatially-varying reflectance of the human face \cite{marschner1999image, ma2007rapid, debevec2000acquiring}.
%{weyrich2006analysis,ghosh2008practical}
%This work, on the other hand, is not focused on modeling the complex spatially-varying reflectance characteristics of the human face (pixel-value distribution). Instead, 

Previous work relied on heavily controlled and calibrated image acquisition, \textit{i.e.}, acquiring multiple samples under controlled illumination or with different polarization, in order to study the fundamentals of reflectance. The acquired data is fit to bidirectional reflectance distribution function (BRDF) and bidirectional surface scattering distribution function (BSSRDF) models \cite{jensen2001practical} to fully represent the spatially-varying reflectance of the human face \cite{ma2007rapid, debevec_acquiring_2000}. We leverage the dichromatic reflection model \cite{shafer_using_1985} for our proposed ``Skin Reflectance Estimate based on Dichromatic Separation'' (SREDS) metric. SREDS generates data-driven, additive, and interpretable measure of skin color.

%a novel, data-driven, computationally inexpensive, and reflection-aware measure to represent the skin color of the subjects derived from traditional RGB face images. 
%Unlike RSR, SREDS does not rely on consistent background and illumination. As a result, it can be easily applied to traditional large scale face datasets without ethnicity labels to reduce the differential performance observed in deep learning based models.

%\subsection{Dichromatic reflection model}
%\label{ssec:Dichromatic}
The Dichromatic Reflection Model (DRM) is a general model for estimating the intrinsic reflections from a standard RGB image. DRM defines two types of reflections, interface (\textit{specular}) and  body (\textit{diffuse}), where each reflection component can be further decomposed into spectral distributions and geometric scaling factors \cite{shafer_using_1985}. The general reflection model proposed in DRM is given in two forms: 
 %klinker_image_1988 
\begin{align}
L (\lambda, i, e,g) &= L_i(\lambda, i, e, g) + L_b(\lambda, i, e, g) 
\label{EQU:DRM_1} \\
& = m_i(i, e, g) c_i (\lambda) + m_b(i, e, g) c_b (\lambda),
\label{EQU:DRM_2}
\end{align}
where in Equation \ref{EQU:DRM_1}, $L_i$, $L_b$, $\lambda$, $i$, $e$, and $g$, respectively, represent radiance of interface reflection, radiance of body reflection, wavelength of light, angle of incidence, angle of existence, and phase angle.  
%The total reflectance proposed in DRM, with $L_i$, $L_b$, $\lambda$, $i$, $e$, and $g$, respectively, represent radiance of interface reflection, radiance of body reflection, wavelength of light, angle of incidence, angle of existence, and phase angle, is given by: 
%
%Equation \ref{EQU:DRM_2} further decomposes 
The factors $L_i$ and $L_b$ are further decomposed into \textit{compositions}, $(c_i , c_b)$, and \textit{magnitudes}, $(m_i , m_b)$ as shown in Equation \ref{EQU:DRM_2}. Composition components (representing shading) depend on the geometry of the object and are independent of the wavelength. Magnitude components represent the wavelength dependent nature of the radiance (representing color) and are independent of the geometry of the object. 

The DRM introduces a set of assumptions to further simplify Equation \ref{EQU:DRM_2} and represent the pixel values in the RGB domain as a tristimulus integration over the amount of received light and sensitivity of the camera at each wavelength. The model assumes that the material of interest is in-homogeneous, opaque, and uniformly colored. Additionally, the material should have one significant specular reflection, isotropic diffuse reflection, and constant spectral distribution across the scene. The proposed tristimulus integration is a linear transformation. As a result, every pixel value in the image can be approximated by a linear combination of interface and body reflection colors of the material as
\begin{equation}
C_L = m_i C_i + m_b C_b,
\label{EQU:DRM_3}
\end{equation}
where $C_L$, $C_i$, $C_b$, $m_i$, and $m_b$, respectively, represent the pixel value, color of interface reflection, color of body reflection, magnitude of interface reflection, and magnitude of body reflection.

%where $C_L$, $C_i$, $C_b$, $m_i$, and $m_b$ respectively represent the pixel value, color of interface reflection, color of body reflection, magnitude of interface reflection and magnitude of body reflection. 

The specular reflection is assumed to have the same spectral power distribution (color) as the incident illumination. The amount of specular reflection relates to the angle of incidence, index of refraction of the material and polarization of the incoming illumination as governed by Fresnel’s laws \cite{shafer_using_1985}, and is largely due to the air-oil interface from the surface of the skin\cite{debevec_acquiring_2000, ghosh_practical_2008}. The diffuse component is generally assumed to be isotropic \cite{shafer_using_1985}, and it is due to the subsurface scattering of light with melanin and hemoglobin components in epidermis and dermis layers of skin \cite{debevec_acquiring_2000, igarashi_appearance_2005}.
%index of refraction of the material and polarization of the incoming illumination as governed by Fresnel’s laws \cite{shafer_using_1985, jenkins1958fundamentals},
%air-oil interface from the surface of the skin\cite{debevec_acquiring_2000, ghosh_practical_2008
%\cite{debevec_acquiring_2000, igarashi_appearance_2005}.

\section{Skin Reflectance Estimate based on Dichromatic Separation (SREDS)}
\label{sec:SREDS}

In this paper, we introduce a new measure of skin color, called the Skin Reflectance Estimate based on Dichromatic Separation (SREDS). %SREDS utilizes a data-driven approach similar to RSR. 
%However, as explained in section \ref{ssec:Relative Skin Reflectance}, while RSR relies on adaptive skin segmentation, requiring consistent lighting, acquisition camera, and background, we propose to estimate the color skin by exploiting the reflection characteristics of the skin instead of relying on consistent background and lightning. In other words, 
SREDS aims to decompose the light reflected off the skin into specular and diffuse components and construct a data-driven skin color metric from the diffuse component of the skin. The diffuse component is due to the interaction of the light with hemoglobin and melanin components in the lower layers of the skin. As a result, it is assumed to be \textit{Lambertian}, \textit{i.e.}, is reflected equally in all directions. Utilizing the diffuse component of the skin allows the SREDS to be insensitive to naturally accruing specular reflections due to uncontrolled illumination and be independent of the angle between the illumination source and the viewer (camera), thus circumventing the requirement for consistent illumination and background as needed to calculate RSR.

%smooth
SREDS extracts patches of skin from the forehead, and right and left cheeks of each face image. Selecting smooth patches of skin allows us to minimize the effect of geometry on the estimated reflection components. Previous work on skin reflectance of human face suggests that cheeks and forehead have similar translucency and isotropic diffuse reflection \cite{weyrich_analysis_2006}. This allow us to rely on the DRM to estimate the diffuse and specular components of the selected skin patches. Given the independence assumption between specular and diffuse components, lack of ground truth, and the smoothness of the selected skin patches (no need to account for the geometry of the face), our problem can be classified as a blind source separation (BSS) problem \cite{comon_handbook_2010}. Independent component analysis (ICA) and principal component analysis (PCA) are two common BSS algorithms \cite{comon_handbook_2010}. Previous work showed the effectiveness of ICA in decomposing melanin and hemoglobin components of skin on samples acquired using high quality cameras in a lab environment %with the help of methyl nicotinate to artificially boost the hemoglobin content of the skin 
\cite{tsumura_image-based_2003}.
\begin{figure}[h]
\begin{minipage}[b]{1\linewidth}
  \centering
  \centerline{\includegraphics[width=6cm]{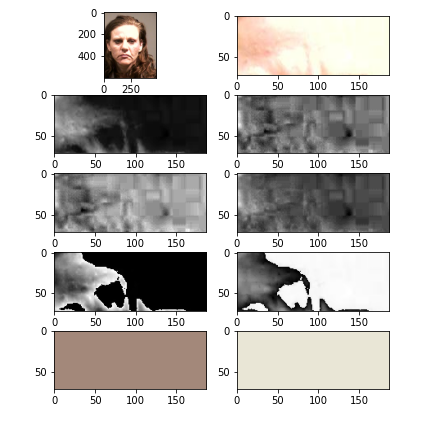}}
\end{minipage}
\caption{Example of a skin patch decomposition. Skin patch of (forehead) (Row-1), PCA (Row-2),  ICA (Row-3), and NNMF (Row-4) decompositions represented as grayscale images (black represents 0 and white represents 1), SREDS Bases (Row-5) - Specular (Right), Diffuse (Left).}
\label{fig:compare}
\end{figure}

%\begin{figure}[h]
%\begin{minipage}[b]{0.3\linewidth}
%  \centering
%  \centerline{\includegraphics[width=3.5cm]{plots/Decomposition_part1.png}}
%\end{minipage}
%\begin{minipage}[b]{0.7\linewidth}
%  \centering
%  \centerline{\includegraphics[width=7cm]{plots/Decomposition_part2.png}}
%\end{minipage} 
%\caption{Example of a skin patch (Forehead) decomposition. Skin Patch (Row-1), PCA (Row-2),  ICA (Row-3), and NNMF (Row-4) decompositions represented as grayscale images (black represents 0 and white represents 1), SREDS Bases (Row-5) - Specular (Left), Diffuse (Right).}
%\label{fig:compare}
%\end{figure}

%We investigated the effectiveness of PCA and ICA algorithms on the selected patches of skin. As there is no reflection ground truth available, we relied on images with high level of visible highlights i.e. specular reflection and manually evaluate the separated components. 
We observed that in a few cases PCA resulted in meaningful but very noisy decomposition. However, ICA completely failed to produce a meaningful decomposition. Figure \ref{fig:compare} represents the PCA (Row-2) and ICA (Row-3) decompositions for a patch of skin from subject's forehead (Row-1). Note that, the sole BSS pipeline does not provide a method for identifying the diffuse component from the decomposed components. Given the ineffectiveness of PCA and ICA decompositions, we propose to relax the independent assumption and utilize the fact that our images and magnitudes ($m_i$, $m_b$) are all strictly positive. We propose to employ Non-negative Matrix Factorization (NMF) \cite{boutsidis_svd_2008} to extract representations of specular and diffuse components from each patch of skin. NMF have shown to perform very well in image based BSS problems \cite{cichocki_fast_2009}.

%cite{cichocki_fast_2009, plaza_quantitative_2004}.
%berry_algorithms_2007
Additionally, unlike PCA, all the weights in NFM are strictly positive, \textit{i.e.}, there is no subtraction involved in representing the input image \cite{lee_learning_1999}. This is in line with the combination of diffuse and specular components proposed in DRM and leads to better interpretability of the decomposed representations \cite{lee_learning_1999}. The NFM decomposition used in SREDS is given by
\begin{equation}
V^t \approx \sum_{i=1}^{M}w_ih^t=WH,
\label{EQU:EQU_NMF}
\end{equation}
where $V^t$ denotes the $t$th skin patch, represented as an $n \times 3$ vector, where $n$ denotes the number of pixels in each skin patch. $W$ and $H$ respectively represent our $n \times 2$ and $2 \times 3$ matrix factors.  
To find the $W$ and $H$ matrices, we solve the following constrained optimization problem: 
\begin{align}
\argmin_{W, H} \quad & \frac{1}{2} ||V^t - WH||^2 \nonumber \\ 
\text{subject to } & W \geq 0 \nonumber \\
& H \geq 0,
\label{EQU:EQU_NMF_2}
\end{align}
where $||.||^2$ denotes the L2 norm. To satisfy the constraints, we employ non-negative double singular value decomposition (NNDSVD) with averaging initialization \cite{boutsidis_svd_2008} to initialize W and H matrices when solving the optimization problem in Equation \ref{EQU:EQU_NMF} numerically. 

%We employ Non-Negative Double Singular Value Decomposition (NNDSVD) with averaging initialization \cite{boutsidis_svd_2008} to initialize W and H matrices. Subsequently, we iteratively minimize our objective function presented in Equation \ref{EQU:EQU_NMF_2} to either reach 200 iterations or stopping tolerance of 5E-3. 

    % \ref{EQU:EQU_NMF_2} represents our objective function where $||.||^2$ denotes the L2 norm. We employ Non-Negative Double Singular Value Decomposition (NNDSVD) with averaging initialization \cite{boutsidis_svd_2008} to initialize W and H matrices. Subsequently, we iteratively minimize our objective function presented in Equation \ref{EQU:EQU_NMF_2} to either reach 200 iterations or stopping tolerance of 5E-3.

As the specular component reflects the color of the illumination and is generally brighter than the color of skin, we select the row of $H$ with higher sum as our specular basis while the other row represents our diffuse basis. 
%Note that the basis assignment process can be replaced with a more sophisticated model by enforcing constrains on the spectral power distribution of the illumination. However, in our experiments, 
We observed that this simple assignment rule can be very effective over multiple datasets. The fourth and the fifth rows of Figure \ref{fig:compare} respectively show the gray scale representation of our NFM decomposition and the RGB representation of the associated specular (right) and diffuse (left) bases. Finally, we utilize kernel principal component analysis (KPCA) \cite{scholkopf_kernel_1997} over our estimated diffuse bases and select the first principal component as our measure of skin color. We evaluated linear, polynomial, and radial basis function (RBF) kernels and observed that the degree-3 polynomial kernel worked best across the evaluated datasets. The final SREDS measure is an average of the three individual measures calculated from forehead and cheeks.  

\section{Region of Interest Selection}
\label{ssec:ROI}

In order to identify regions of interest (ROI) to compute the skin color metrics, we use facial landmark detection. Given an input image, a facial landmark predictor attempts to identify key points of interest based on the shape of the face using 
a two-step process. First, a face is detected in an image. Then, key facial landmarks are detected. In this paper, we use the Dlib library to detect key facial landmarks \cite{kazemi2014one}. 
%In order to identify regions of interest (ROI) to compute the skin color metrics, we use facial landmark detection. Given an input image, a facial landmark predictor attempts identify key points of interest. In this paper, we use the Dlib library to detect key facial landmarks \cite{kazemi2014one}.
%In our work, we detect and extract important regions on the face using shape prediction methods. Detecting facial landmarks is 
%a two-step process: first, a face is detected in an image. Then, key facial landmarks are detected. 
% The facial landmark detector in Dlib is trained on a set of images with pre-labeled facial landmarks and the probability of the distance between pairs of input pixels. These images are manually labeled with the coordinates of regions surrounding each facial structure. With this training set, an ensemble of regression trees are trained to estimate the facial landmark positions directly from pixel intensities to generate a detector. 
Dlib uses this pre-trained facial landmark detector to estimate the location of 68 coordinates that map to facial structures. Then using the landmarks near cheeks and the forehead, we extract rectangular patches from the right cheek, left cheek, forehead, and background for all subjects. The forehead patch is extracted by using the distance between the eye landmarks. The cheek landmarks extracted by estimating the distance between jawline and lips. An example of the extracted crops are shown in Figure \ref{fig:ASS}.

\section{Datasets}
\label{ssec:dataset}

\begin{table*}[!ht]
\caption{Intra-subject variability over multiple datasets. S1: Number of Subjects, S2: Average number of samples per subject, S3: Total number of samples (after cleaning), ITA:Individual Typology Angle, RSR: Relative Skin Reflectance, SREDS: Skin Reflectance Estimate based on Dichromatic Separation, RSR*:Relative Skin Reflectance using skin patches instead of adaptive segmentation.}  
\centering
\begin{tabular}{|p{3.3cm}|| p{1cm}| p{1cm}| p{1cm}| p{1cm}| p{1cm}| p{1cm}| p{1cm}|} 
\hline
Dataset & ITA & RSR & SREDS & RSR* & \# S1 & \# S2 & \# S3\\
\hline
\hline
Multi-PIE - High Resolution & 0.401 & 0.307 & \textcolor{magenta}{$\boldsymbol{0.138}$} & 0.356 & 314 & 3.6 & 1,170 \\
\hline
Multi-PIE - Multi View & 0.926 & 0.860 & \textcolor{magenta} {$\boldsymbol{0.820}$} & 0.881 & 314& 447 & 150,668 \\
\hline
MEDS-II & \textcolor{magenta} {$\boldsymbol{0.448}$} & 0.493 & 0.463 & 0.475 & 171 & 3.3 & 834 \\
\hline
Morph-II & 0.645 & 0.539 & \textcolor{magenta} {$\boldsymbol{0.419}$} & 0.562 & 13,152 & 4.1 & 55,045 \\
\hline
\end{tabular}
\label{table:ComparisonT}
\end{table*}

%We have used the Multi-PIE Multi-view dataset \cite{sim2002cmu}, the Multiple Encounter Dataset (MEDS-II) \cite{founds2011nist}, and the Morph dataset \cite{ricanek2006morph} for this investigation. 

\subsubsection{Multi-PIE}
\label{sssec:Multipie}
The CMU Multi-PIE face database \cite{sim2002cmu} contains more than 750,000 images of 337 people recorded in up to four sessions over five months. Subjects were imaged under 15 viewpoints and 19 illumination conditions while displaying a range of facial expressions. High-resolution frontal images were acquired as well. We selected three viewpoints ($14\_0$, $05\_1$, $05\_0$) where full views of the face were captured and images where the facial landmark detection failed are removed. Overall, we select 150,668 images out of 750,000 images from 314 subjects.

%We selected three viewpoints out of 15, where full views of the face were captured. Overall, we select 150,668 images out of 750,000 images from 314 subjects. 

\subsubsection{MEDS-II}
\label{sssec:MEDS}

Multiple Encounter Dataset (MEDS-II)  \cite{founds2011nist} by NIST is a data corpus curated from a selection of deceased subjects with prior multiple encounters. It consists of 1170 images from 425 subjects. We utilized the 856 mugshot images for our research. Then, images where facial landmark detection failed are removed, resulting in a reduced image set consisting of 836 images from 171 subjects.
%Out of 1170 images, we used 856 mugshot images for our research.
\subsubsection{Morph-II}
\label{sssec:Morph}
The academic MORPH database is a non-commercial dataset collected over 5 years with multiple images of each subject (longitudinal). It is not a controlled collection (\textit{i.e.}, it was curated in real-world conditions). This dataset contains 55,139 unique images of more than 13,000 individuals, spanning from 2003 to late 2007. Ages of the subjects range from 16 to 77, with a median age of 33. The average number of images per individual is 4. The average time between photos is 164 days. Images where the facial landmark detection failed are removed, resulting in a reduced image set consisting of 55,063 images from 13,152 subjects.

\section{Results and Discussion}
\label{sec:ResDis}
%Goals
    % We need to show that this is an interpretable measure
    % We need to show that the measure is stable (low intra-subject variability)
    % We need to show that the measure is meaningful (follows race labels)
    % Show that SREDs continuously flows from light skin to dark skin according to our labeled datasets

% Tell the story of our analysis when we....
    % Comparison in high-res images
    % Comparison under illumination
    % What happens with poor quality images
    % When we add ethnicity information to evaluate the performance of the estimator
    % What do the NNMF decompositions tell us about the results
%We have used the Multi-PIE Multi-view dataset \cite{sim2002cmu}, the Multiple Encounter Dataset (MEDS-II) \cite{founds2011nist}, and the Morph dataset \cite{ricanek2006morph} for this investigation.

In this work, we evaluate and compare the performance of ITA, RSR, and SREDS under both controlled (Multi-PIE \cite{sim2002cmu}) and uncontrolled illumination (MEDS-II \cite{founds2011nist}, Morph-II \cite{ricanek2006morph}). There is no objective measure of ground truth without a direct measure of skin reflectance controlling for all factors.  However, we propose that the hypothetical perfect skin color metric has to be stable, \textit{i.e} robust against all possible variations such as changes in illumination or camera sensitivity; thus, it is highly desirable for any skin color metric to have lower intra-subject variability. In order to fairly compare the evaluated metrics, we individually normalize each metric to have zero mean and unit variance and use the intra-subject variability (standard deviation) of each metric to evaluate its effectiveness. Table \ref{table:ComparisonT} presents the average intra-subject variability of the ITA, RSR, and SREDS over the evaluated datasets. 

%The lack of a subjective measure of a skin color prevents us from evaluating using traditional supervised learning metrics.
%hinders the comparison of the evaluated metrics using traditional classification/regression metrics. 
%Our results suggest that %SREDS provides more stable (lower intra-subject variability) across all three datasets while not relying on consistent background or illumination.   
%also providing a continues and interpretable progression from darker to lighter skin colors.
%Additionally, we investigated the relationship between each measure and the ethnicity labels provided in MEDS-II and Morph-II detests. Figure \ref{fig:MEDS_Ethnicity} depicts the distribution of black and white subjects over each measure.

%The image acquisition environment of Multi-PIE dataset allows us to isolate the effect of illumination change on the intra-subject variability of the evaluated metrics, as it includes both a constant background and controlled variation of illumination. 

%Subsequently, we utilize the multi-view section of the dataset and evaluate the sensitivity of the measures to illumination change. 

\begin{figure}[!ht]
\begin{minipage}[b]{1.0\linewidth}
  \centering
  \centerline{\includegraphics[width=8.6cm]{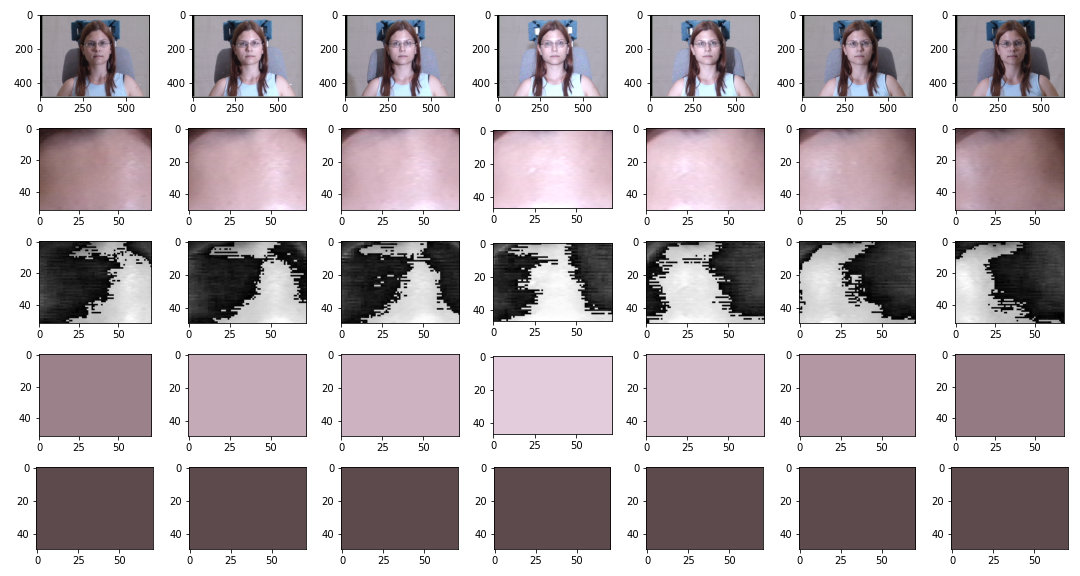}}
\end{minipage}
\caption{Diffuse and specular components vs. illumination change in Multi-PIE dataset. Row-1: face images, Row-2: forehead skin patches, Row-3: weights associated with the specular basis, Row-4: specular bases, Row-5: diffuse bases. Best viewed in color.}
\label{fig:illum}
\end{figure}

The image acquisition environment of Multi-PIE dataset provides us with both a constant background and controlled variation of illumination. This allow us to isolate the effect of illumination change on the intra-subject variability of the evaluated metrics. Our results suggests that SREDS outperforms the other two algorithms in both controlled and varying illumination environments and without relying on the constant background. Note that all measures show higher intra-subject variability under varying illumination. Figure \ref{fig:illum} illustrates SREDS over multiple samples of the same subject collected while changing the angle incidence from ${-45\si{\degree}}$ (most left) to ${45\si{\degree}}$ (most right) with step size of ${15\si{\degree}}$. SREDS is able to extract a stable diffuse component while the specular component is brighter with the increase of the angle of incident. 

%Not that under varying illumination, all three measure show higher intra-subject variability compared to the high-resolution captures. 
%Our result suggests that SREDS outperforms both RSR and ITA without relying on the constant background. 

%Note that, even-though RSR is designed for constant illumination it outperforms ITA while operating under illumination change. This is in line with previous observation that suggest the sensitively of ITA to illumination change.
%As we can see,  

%Figure \ref{fig:illum} illustrates the subjects image(row-1), forehead skin patches (row-2), weights associated with the specular basis (row-3), specular (row-4) and diffuse (row-5) bases of SREDS. As we can see,  

\begin{figure}[ht]
\begin{minipage}[b]{1\linewidth}
  \centering
  \centerline{\includegraphics[height=2.5cm]{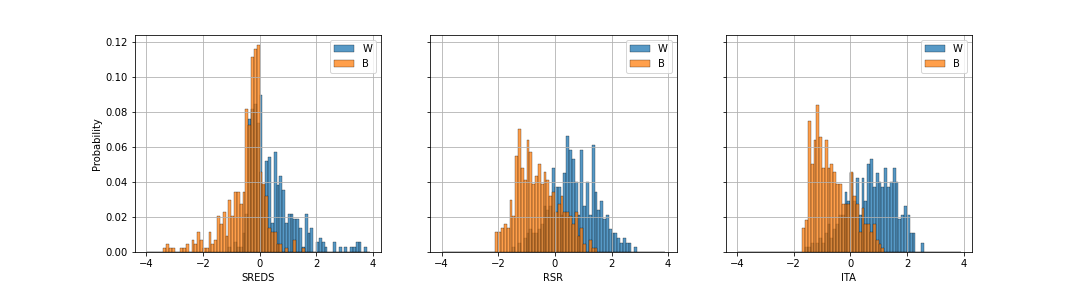}}
  \centerline{(a) MEDS-II}\medskip
\end{minipage}
\begin{minipage}[b]{1\linewidth}
  \centering
  \centerline{\includegraphics[height=2.5cm]{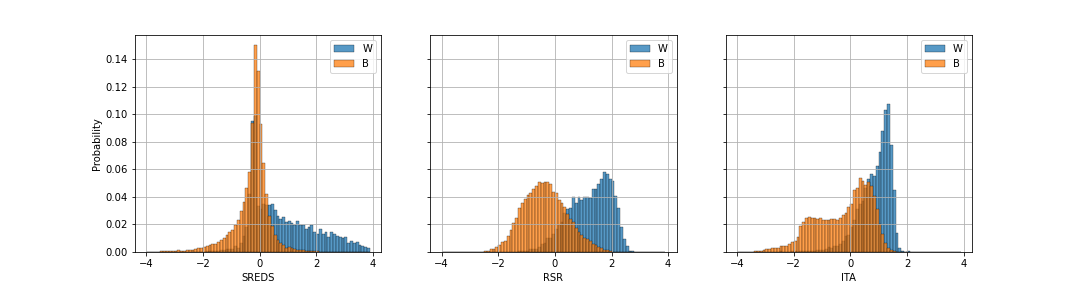}}
  \centerline{(b) Morph-II}\medskip
\end{minipage}
\caption{Histograms of SREDS, RSR and ITA for black (B) and white (W) subjects in MEDS-II and Morph-II datasets. Best viewed in color.}
\label{fig:MEDS_Ethnicity}
\end{figure}

We also evaluated all three algorithms using MEDS-II and Morph-II datasets (uncontrolled illumination). SREDS outperforms RSR and ITA in the larger and wider Morph-II dataset. However, we also observe that ITA provides marginally better performance then SREDS and RSR in the MEDS-II dataset. We suspect this might be due to the smaller size of the MEDS-II and the data-driven nature of the RSR and SREDS. Additionally, we investigated the relationship between each measure and the ethnicity labels provided in the datasets. Figure \ref{fig:MEDS_Ethnicity} depicts the distribution of black and white subjects over each measure. The distribution of SREDS measurements over black and white subjects in uncontrolled illumination suggests that SREDS provides a meaningful progression from darker to lighter skin tone subjects without relying on consistent background, illumination or camera sensitivity. 

\subsection{Limitations and Future work}
Our results suggests that SREDS can produce a continuous and interpretable skin color metric across different image acquisition environments and without relying on constant background or illumination. However, similar to the other evaluated measures, SREDS is still susceptible to illumination change. In future work we aim to improve upon this work by integrating a more complex basis assignment process. Our current model does not utilize the wavelength dependent nature of reflection in the assignment process and treats Red, Green, and Blue channels equally. Utilizing this fact can potentially improve the bases assignment process. Finally, in future work we plan to evaluate the correlation between the estimated SREDS values and direct measurements of skin reflectance using a DSM III skin colormeter.

\section{Conclusions}
\label{sec:conclusion}

While there is currently a large amount of active research providing solutions for removing bias from facial recognition systems, many of the methodologies assume the existence of large-scale data that is labeled and in-domain. Operational systems rarely can fulfil this requirement. These restrictions make the solutions very difficult to apply outside of the structured and labeled datasets generally used to benchmark algorithms. 
%In this research, we targeted this problem
Inspired by this problem, we developed an interpretable skin tone estimate with few restrictions that can provide performance comparable or superior to similar methods in the literature. This estimate will provide a way for facial recognition applications without access to large demographically labeled datasets within their domain to make effective use of methods to reduce difference performance and promote fairness in face recognition.

%\bibliographystyle{IEEEbib}{9}
%\bibliography{refs}

\bibliographystyle{IEEEtran}
\bibliography{IEEEabrv,refs}
\end{document}